\documentclass[%
 reprint,
 amsmath,amssymb,
 aps,
pra,
]{revtex4-1}

\usepackage{graphicx}
\usepackage{dcolumn}
\usepackage{bm}

\begin{document}

\title{An eavesdropping attack on a trusted continuous-variable quantum random number generator}

\author{Johannes Thewes}
\author{Carolin L\"uders}%
\author{Marc A{\ss}mann}
 \email{marc.assmann@tu-dortmund.de}
\affiliation{%
 Experimentelle Physik 2,
	Technische Universit\"at Dortmund,
	D-44221 Dortmund, Germany
}%

\date{\today}

\begin{abstract}
Harnessing quantum processes is an efficient method to generate truly indeterministic random numbers, which are of fundamental importance for cryptographic protocols, security applications or Monte-Carlo simulations. Recently, quantum random number generators based on continuous variables have gathered a lot of attention due to the potentially high bit rates they can deliver. Especially quadrature measurements on shot-noise limited states have been studied in detail as they do not offer any side information to potential adversaries under ideal experimental conditions. However, they may be subject to additional classical noise beyond the quantum limit, which may become a source of side information for eavesdroppers. While such eavesdropping attacks have been investigated in theory in some detail, experimental studies are still rare. We experimentally realize a continuous variable eavesdropping attack, based on heterodyne detection, on a trusted quantum random number generator and discuss the limitations for secure random number generation that arise.
\end{abstract}

\maketitle


\section{Introduction}
Randomness is a fundamental resource for many technological applications ranging from cryptography \cite{Shannon1949} and computer simulations \cite{Metropolis1949} up to gaming. For those purposes which only require uniformly distributed numbers and do not focus on security and full unpredictability, deterministic approaches may be used. This includes pseudorandom number generators that generate a deterministic sequence of random numbers based on a seed or classical random number generators that exploit resources such as classical noise to create random numbers that are in principle deterministic, but in practice hard to predict \cite{Uchida2008,Kanter2009}. However, it is not possible to strictly validate the unpredictability of such random numbers. Accordingly, if security is the highest priority, quantum random number generators (QRNGs) based on inherently unpredictable quantum processes are considered as the best choice for random number generation \cite{Calude2010}. Plenty of physical realizations of QRNGs have been demonstrated based on effects ranging from phase fluctuations \cite{Qi2010} and photon arrival times \cite{Wahl2011,Ma2005,Fuerst2010} over polarization fluctuations \cite{Fiorentino2007}, which-way information at beams splitters \cite{Stefanov2000, Jennewein2000} and spontaneous emission noise \cite{Stipcevic2007,Williams2010} up to photon number noise detection using mobile phone cameras \cite{Sanguinetti2014}.
\\While most of these implementations rely on discrete variables, recently also realizations based on continuous variables such as the quadratures of the light field form an attractive alternative as they commonly offer fast random number generation rates \cite{Avesani2018} and several bits of entropy per measurement \cite{Gabriel2010}. Determining the exact amount of random bits that can be extracted from a single measurement is a nontrivial task for any random number generator as it depends on the amount of quantum side information that may be available to potential eavesdroppers. Accordingly, the extractable randomness depends on the implementation of the QRNG, the application and on whether several parts of the QRNG, such as the source or the detector can be considered as trusted. In general, several groups of QRNGs can be identified \cite{Ma2016}. Most practical QRNGs, including most of those available commercially belong to the category of trusted QRNGs. These are known to deliver secure true random numbers if the physical implementation of the device is faithful to the model assumed. However, if these assumptions are not fulfilled, for example in the case of a malicious manufacturer manipulating either the source or the readout device, the output may seem to be genuinely random to the user, although it is in fact known to some adversary, so the random numbers may not be private. A precreated string of random numbers copied to the device by some adversary is a prime example for such a scenario. In this case, there is no way to certify whether a minimal amount of genuine randomness is present or not. Still, such designs show enduring popularity due to relatively low cost and ease of implementation as well as high bit rates.\\
On the other hand, device-independent implementations of QRNGs based on entanglement can be built, which are in fact self-testing and indeed deliver private random numbers \cite{Pironio2010,Bierhorst2018}, so that a verifiable amount of randomness can be extracted without a need to trust the implementation. However, these device-independent protocols require non-locality, which results in complex setups and rather low bit rates up to about 200 bits per second \cite{Liu2018}.\\

Semi-device-independent QRNGs are an attempt to merge the advantages of both classes of QRNGs. They still provide high bit rates up to several gigabit per second \cite{Avesani2018} and offer a certain certifiable amount of randomness, but require weaker assumptions with respect to the faithfulness of the QRNG. Possible protocols include QRNGs with a trusted readout section, but an untrusted source \cite{Cao2016}, the opposite case \cite{Cao2015}, assumptions on the overlap of states in phase space \cite{Brask2017} or the size of the underlying Hilbert space \cite{Lunghi2015}.\\

In the cases of fully trusted and device-independent QRNGs, it is possible to determine the actual amount of randomness per measurement delivered by the device because the available quantum side information is well defined either by the definition of the protocol or the self-testing process. Here, theoretical lower bounds for the extractable randomness are well established. On the contrary, for semi-device-independent realizations of QRNGs, the exact amount of quantum side information available to potential eavesdroppers depends strongly on the actual protocol and assumptions used and is not necessarily easy to determine theoretically. Here, experimental studies of actual eavesdropping attacks may prove very helpful to adequately quantify the quantum side information present, but are not necessarily easy to conduct \cite{Huang2013,Huang2014,Qin2018,Shihan2016}. In the following, we present the experimental realization of such an eavesdropping attack on a trusted continuous variable QRNG based on homodyne detection of thermal light \cite{Qi2017}. In a fully trusted scenario, where the attacker has no access to the source and the light fields, these QRNGs are known to offer a significant enhancement of the extractable randomness compared to QRNGs based on homodyne detection of the vacuum state. We find that in a scenario of reduced trust, where the state of the light field is still assured to be thermal, but an adversary may gain access to the light field using beam splitters, an eavesdropping attack on such a trusted device may reduce the extractable randomness from a thermal state down to the amount that may be extracted from a vacuum state.

\section{The Continuous Variable QRNG}
We investigate a continuous variable QRNG. The QRNG user (Alice) mixes a light field of interest with a strong single mode local oscillator (LO) on a beam splitter and guides it to a homodyne detector, which performs differential photodetection and yields a voltage signal proportional to one quadrature of the electromagnetic field \cite{Jakeman1975,Yuen1983}. The measured quadrature $\tilde{X}$ is then equivalent to a random value drawn from a probability distribution given by the integral projection $W_{\tilde{X}}$ of the Wigner function of the light field along a direction determined by the relative phase $\varphi$ between the signal and Alice's local oscillator. As the relative phase between the LO and the signal is not fixed, we define the quadrature measured by Alice as $\tilde{X}$ and the orthogonal quadrature which is unknown to Alice as $\tilde{P}$. The measured voltage signal is then digitized using an analog-to-digital converter and effectively sorted into bins of finite width. Subsequently, the index of these bins acts as the random number. In the following, we will denote the binned quadratures as $X$ and $P$. Such QRNGs have already been realized using the vacuum \cite{Marangon2017} or thermal light \cite{Qi2017} as input states. The random numbers determined this way are usually not distributed uniformly, so it is possible to perform additional randomness extraction \cite{Ma2013} to create a compressed shorter string composed of independent identically distributed random numbers.\\

\begin{figure*}
\includegraphics[width=0.95\textwidth]{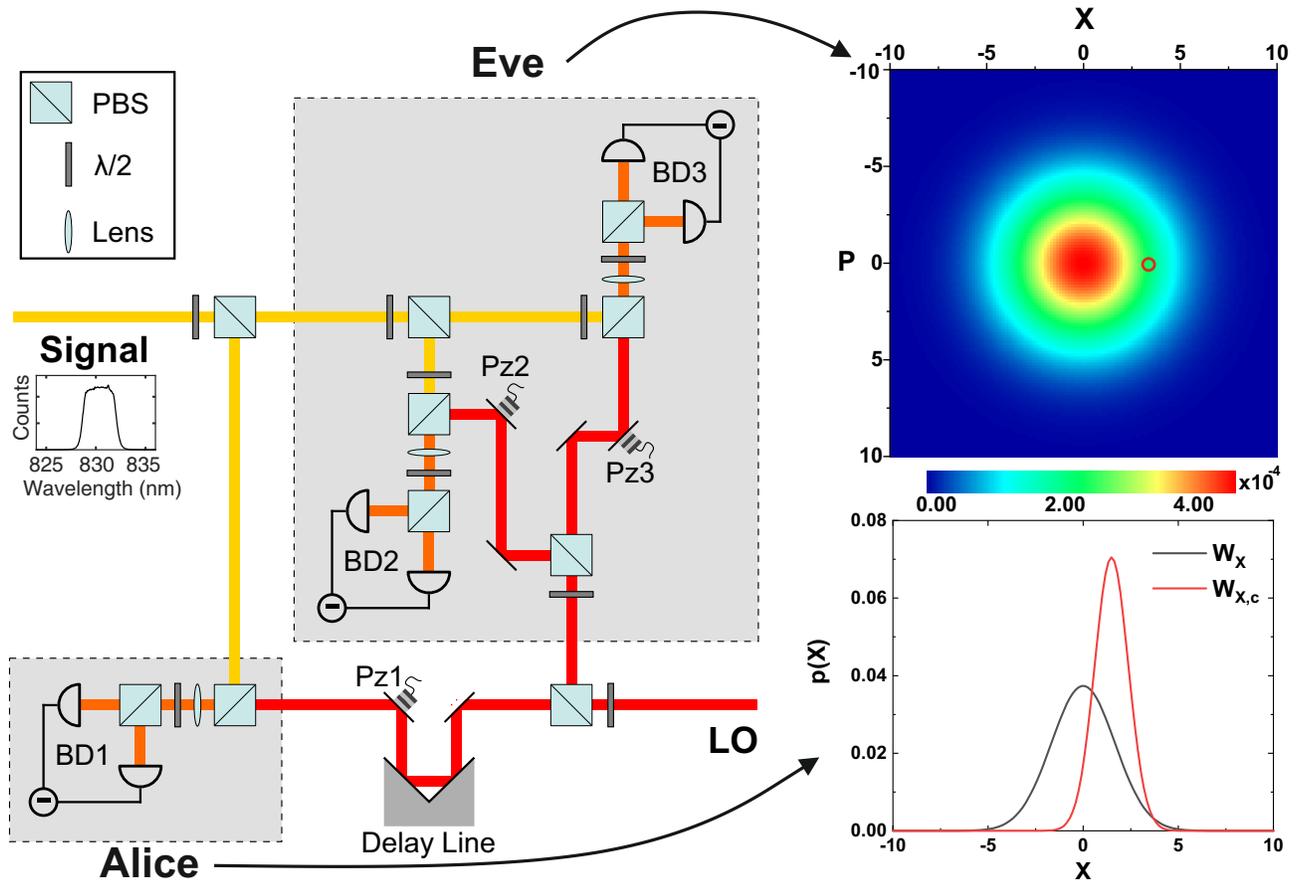}
\caption{\label{fig:Setup} Experimental setup for the eavesdropping attack. Alice receives the light field of interest, superposes it with a strong pulsed local oscillator at a certain time determined by the position of a delay line and performs homodyne detection using balanced detector BD1. Every detected quadrature is a random number occuring with relative frequencies given by the projection $W_X$ of the Wigner function of Alice's signal along one axis chosen by the position of the phase shifting piezo element Pz1. Eve redirects an adjustable fraction of the signal and also a part of the local oscillator to her part of the setup and performs heterodyne detection. The settings of both the delay line and Pz1 are known to Eve, so she samples the same section of the signal Alice does and she identifies the relative phase of her local oscillator beams compared to Alice's local oscillator. Depending on the pair of quadratures measured, Eve then constructs the conditional Husimi function for the signal seen by Alice, reconstructs the conditional Wigner function $W_c$ for this signal and determines its projection $W_{X,c}$ along the axis used by Alice. An exemplary set of experimental data is shown in the right panels. The upper panel shows the Husimi function in Eve's channel for $n_{\text{Eve}}=7.09$. The small red ring denotes a measured quadrature set of $X_E=3,P_E=0$, which results in the $W_{X,c}$ denoted by the red line in the lower panel for $n_{\text{Alice}}=2.28$ and PZ1 set such that Alice measures the $X$ quadrature. $W_{X,c}$ yields a significantly narrower probability distribution compared to $W_X$ given by the black line, indicating the additional side information gathered by Eve.}
\end{figure*}

Here, we implement a continuous variable eavesdropping attack by adding a half-wave plate and a polarizing beam splitter to redirect some of the signal towards a second detection channel, where a malicious eavesdropper (Eve) placed two additional homodyne detectors and performs heterodyne detection, which allows her to determine the Husimi Q-function of the redirected light field. To this end, Eve uses local oscillator beams with orthogonal phase, which are derived from Alice's local oscillator beam and thus share a fixed phase relationship with it. The result of this measurement provides Eve with some side information $\varepsilon$ about the random number generated by Alice. The full setup is depicted in figure \ref{fig:Setup} and implemented as follows:\\
We use a pulsed Ti:sapphire laser delivering pulses with a duration of about 120\,fs at a repetition rate of about 75.4\,MHz as the local oscillator and a Toptica DL pro continuous wave diode laser operated far below threshold as the thermal light source for the QRNG. The latter has no fixed phase relative to the local oscillator. Both are set to a central wavelength of 830\,nm. The homodyne detectors were provided by Femto (model HCA-S), are identical to the ones used in \cite{Lueders2018} and feature a bandwidth of 100\,MHz and a transimpedance gain of 5\,kV/A. The signals from each detector were filtered using band-stop filters at 75.4 and 150.8\,MHz and a low-pass filter at 100\,MHz to remove all harmonics of the laser repetition rate. Afterwards the signals were amplified by a factor of 5 using a 300\,MHz SR455 Voltage amplifier from Stanford Research Systems and digitized using an 8-bit 5\,Gs/s analog-to-digital converter M4i.2234-x8 from spectrum, which provides a sampling rate of 1.25\,Gs/s per detection channel. It is important to note that the multi-channel detection setup does not only measure a histogram of quadratures, but records every single detected quadrature with a timestamp, which allows us to determine the additional amount of information gained by Eve a posteriori. The detector is shot noise-limited for LO powers above 1\,mW and yields a common-mode rejection ratio of 68.9\,dB and a low-frequency shot-noise clearance of about 12.5\,dB for the LO powers used here. For calibration purposes, we perform a test measurement using only the LO in the absence of any signal, which corresponds to a measurement of the vacuum state. As the quadrature variance of the vacuum state is known, this procedure yields a conversion factor between the measured voltages and the quadratures of the light field. The electronic noise present broadens the vacuum state variance by about 3$\%$. As this is small compared to our chosen quadrature bin width, we will neglect this effect. Also, we would like to point out that each individual recorded voltage is digitized with 8-bit resolution, but the signal of interest is the total voltage time-integrated over a whole local oscillator pulse, which may effectively result in a higher resolution. Still, we chose to discretize all the quadratures used in this manuscript with 8 bit resolution, corresponding to a bin width of 0.15625 for $X$ and $P$. An additional advantage of using a pulsed LO lies in the high temporal resolution it allows us to achieve. As the instantaneous amplitude and phase of the thermal light field will typically change randomly on a timescale given by the sub-ps coherence time of the light field, using a short pulsed LO ensures that we do not average the signal over several coherence times. Throughout this manuscript, we assume that Eve has full access to the calibration procedures of all detectors. Therefore, our attack targets both the detectors and the light fields involved. Also, due to the high laser repetition rates we use, we can already perform millions of experimental runs in a few tens of milliseconds. On this timescale, mechanical drifts of the relative phase of the different LOs derived from the same laser are not very significant, so we do not perform any active phase stabilization. As we assume that Eve has access to the calibration procedures, this is sufficient information for her to align her LOs phases with respect to the settings chosen by Alice. For a real long-term eavesdropping attack, Eve would have to compensate long term phase drifts or timing errors as well. If the timing used by Eve is off by more than approximately one coherence time of the signal with respect to the timing chosen by Alice, the side information available to Eve will be reduced significantly.\\

\section{Quantum Side Information}
In this section, we will discuss the amount of randomness present and several measures that quantify it both for the fully trusted QRNG scenario and for a semi-device-independent case, where the source itself is trusted, but an eavesdropping attack on the signal is possible in the readout section. First, we will introduce the trusted QRNG case and then discuss how quantum side information reduces the extractable randomness. To this end, we will discuss two different measures for the extractable randomness. First, we will discuss the min-entropy, which is the relevant figure of merit for determining the secure generation rate of a QRNG in the case of an attack. Second, we will discuss the expected guesswork, which is the quantity of interest in the case of a brute force attack, where Eve is allowed to make several guesses. It is not relevant for the security of QRNGs, but useful for other fields, such as quantum state estimation.

\subsection{Min-entropy}
In order to evaluate the effective number of random bits Alice can extract per measurement, she needs to evaluate the amount of knowledge about her measurement result available to Eve, which defines the remaining amount of true randomness left. In the case of a trusted QRNG, the amount of true randomness is directly linked to the min-entropy of $X$ given by:
\begin{equation}
\label{eq:HMin}
H_{\text{min}}(X)=-\log_2 \big(\text{max}(W_X)\big),
\end{equation}
where $W_X$ corresponds to the binned Wigner function using the bin size chosen for $X$ and $P$. The min-entropy yields the effective number of random bits per measurement and is equivalent to assuming that Eve makes a single guess on the outcome of the measurement and chooses the bin with highest probability. Thus, the min-entropy defines a solid lower bound for the number of extractable random bits $\beta_{\text{s}}$ that is also easy to treat theoretically. For a fixed experimental resolution it depends only on the largest value of the probability distribution, but not on its precise shape.\\
In the presence of quantum side information $\varepsilon$ available to Eve, evaluating the extractable randomness becomes more sophisticated. The Leftover Hash Lemma \cite{Tomamichel2011} states that in the presence of quantum side information, the amount of extractable randomness is given by the quantum conditional min-entropy \cite{Koenig2009}
\begin{equation}
\label{eq:CondHMin}
H_{\text{min}}(X|\varepsilon)=-\log_2 \big(\text{max}(W_{X,c})\big),
\end{equation}
where $W_{X,c}$ denotes the integral projection of the conditional Wigner function at Alice's detector depending on the quantum side information available to Eve. As $\text{max}(W_{X,c}) > \text{max}(W_X)$, the effective number of random bits per measurement decreases correspondingly.

\subsection{Expected Guesswork}
The min-entropy focuses on the probability of Eve guessing the correct experimental outcome on her first try. In contrast, one may also assume a brute force attack on the random number generator, where Eve is allowed to make several consecutive guesses. This scenario is obviously not relevant with respect to the security of QRNGs as no realistic protocol will allow for brute force attacks, but it is of interest for applications such as quantum state tomography of time-varying fields, where an experimenter may for example be interested in identifying the state of a light field that changes slowly with time through repeated measurements. Here, one is rather interested in finding the value that has the highest likelihood of being in accordance with all experimental results. Accordingly, it is mandatory to consider also the possible experimental outcomes beyond the one with highest probability in order to determine the leftover randomness or uncertainty in such a scenario.\\
In the case that Eve is allowed to make several guesses and perform a brute force attack, it is harder to quantify the precise amount of randomness present adequately as the whole probability distribution needs to be taken into account. A reasonable quantity to characterize the amount of randomness present in such a scenario is the expected guesswork 
\begin{equation}
\langle G(X) \rangle = \sum_x G(x) p(X=x),
\end{equation}
which corresponds to the average number of guesses Eve needs to make to find the correct result. Here, $G(x)$ represents the number of guesses required to find the correct result $X$ in the case that $X=x$. The expected guesswork depends strongly on the strategy used by Eve. Assuming that there are $n$ different possible outcomes, Eve may apply an optimized attack pattern, where she orders her guesses from highest to lowest probability of success, so that $G(x_{max})=1$ and $G(x_{min})=n$, where $x_{max}$ and $x_{min}$ denote the outcomes with highest and lowest probability, respectively. Naively, one might expect that this number is related to the Shannon entropy of $X$ and it is well known that a lower bound for $\langle G(X) \rangle$ exists, which is directly related to Shannon entropy. However, this bound is not tight, there is no nontrivial upper bound and in most cases $\langle G(X) \rangle$ bears little relation to it \cite{Massey1994}. Instead, it has been pointed out \cite{Arikan1996} that for a series of $k$ measurements with outcomes that Eve tries to guess, the asymptotic moments of the expected guesswork for guessing all of the outcomes correctly simultaneously are directly related to the specific R\'{e}nyi entropy of $X$ as follows:
\begin{equation}
\lim_{k\rightarrow\infty}\frac{1}{k}\log_2 \langle G(X_k)^\alpha\rangle=(1+\alpha)\log_2\sum_x p(X=x)^\frac{1}{1+\alpha}.
\end{equation}
However, this quantity is meaningful only for large $k$. For estimating how prone to a brute force attack a QRNG is, $\langle G(X) \rangle$ itself is indeed a more suitable indicator in most cases. For independent identically distributed random numbers consisting of $\beta$ random bits, a brute force attacker attempting to guess the random number will require on average a number of guesses given by:
\begin{equation}
\langle G(X)_{iid} \rangle=2^{\beta-1}+0.5.
\label{eq:guesses}
\end{equation}
As the random number distribution in our scenario follows the Gaussian distribution of quadratures, it is not uniform prior to randomness extraction. Eve may therefore perform a modified optimal brute force attack based on educated guesses and it is expected that $\langle G(X)\rangle\leq\langle G(X)_{iid} \rangle$.\\
In the presence of side information, also the expected guesswork must be replaced by the conditional expected guesswork:
\begin{equation}
\langle G(X|\varepsilon) \rangle=\sum_x G_{\varepsilon}(x) p(X=x|\varepsilon),
\end{equation}
where Eve reorders her guesses according to the side information available.

\section{The Eavesdropping Attack}
We first discuss how Eve gains access to $\varepsilon$. At the beam splitter that Eve uses for the attack, the thermal signal field $X_{th},P_{th}$ enters at one input port and a vacuum state $X_{vac},P_{vac}$ enters at the other. Both become mixed and the total output may be described by a joint Husimi-Q function, which depends on the following effective mixed quadratures, where $X_A,P_A$ and $X_E,P_E$ describe the values for Alice's and Eve's part of the setup, respectively:
\begin{align}
\label{eq:QuadDef}
X_{th}=&r X_A - t X_E\\
P_{th}=&r P_A - t P_E\nonumber\\
X_{vac}=&t X_A + r X_E\nonumber\\
P_{vac}=&t P_A + r P_E.\nonumber
\end{align}
Here, $r$ and $t$ denote the reflection and transmission coefficients of the beam splitter. In these variables, the joint Husimi-Q function is given by:
\begin{align}
\label{eq:CondHus}
H_j(X_1,P_1,X_2,P_2)=&\frac{H^4_{res}}{4\pi^2 (n+1)}\exp{(-\frac{1}{2} \frac{X_{th}^2 +P_{th}^2}{n+1})}\\
&\times\exp{(-\frac{1}{2}(X_{vac}^2+P_{vac}^2))}, \nonumber
\end{align}
where $n$ is the total signal photon number and $H_{res}$ represents the effective resolution due to the quadrature bin size. As Eve has access to the signal photon number, for every single pair of quadratures $X_E$,$P_E$ she measures, she may insert these values into Eq.(\ref{eq:QuadDef}) and (\ref{eq:CondHus}). This reduces the joint Husimi-Q function $H_j$ to a conditional Husimi-Q function $H_c$ for the expected outcome at Alice's part of the setup for every measurement Eve performs. For signal light fields that are not minimum-uncertainty states, $H_c$ will be a narrower distribution than the Husimi-Q function of the bare signal state, so Eve has gained some quantum side information about the instantaneous state of the light field. In order to predict the value Alice will measure, Eve now needs to convert this conditional Husimi-Q function into a conditional Wigner function $W_c$. As $H_c$ essentially is the convolution of $W_c$ with a Gaussian corresponding to the vacuum state, this requires a deconvolution which can be cumbersome in general. However, in our scenario the range of possible states of the signal light field is limited to well-behaved classical light fields, which are Gaussian states themselves. In this case, Eve can simply construct $W_c$ from $H_c$ using Gaussian deconvolution, which corresponds to constructing a Gaussian with the same mean quadrature values $\langle X\rangle$ and $\langle P\rangle$, but standard deviations $\sigma_X$ and $\sigma_P$ reduced by 0.5. We assume that Eve also has access to the settings of the phase shifting element used by Alice, so she can also perform a projection of the determined conditional Wigner function along the axis chosen by Alice. Eve may then sort the possible random number bins from highest to lowest probability and begin guessing.\\

After repeating this experiment 2 million times, we keep track of each individual result measured by Alice and the number of guesses Eve needed to make to guess the correct value measured by Alice in each experimental run. The average number of guesses Eve needed to make now directly provides an experimentally determined value the expected guesswork and the relative frequency of Eve being correct on the first guess allows one to determine $H_{min}$ by inserting this relative frequency instead of $\text{max}(W_X)$ and $\text{max}(W_{X,c})$ in equations (\ref{eq:HMin}) and (\ref{eq:CondHMin}), respectively. Comparing both quantities for the conditional case in the presence of quantum side information and the unconditional case provides information about how successful the eavesdropping attack actually is. We would further like to point out that this experimentally determined value automatically includes influences such as electronic noise, and potential mismatch of the chosen discretization values or calibration procedures.\\

\section{QRNG security}
\begin{figure}
\includegraphics[width=0.95\columnwidth]{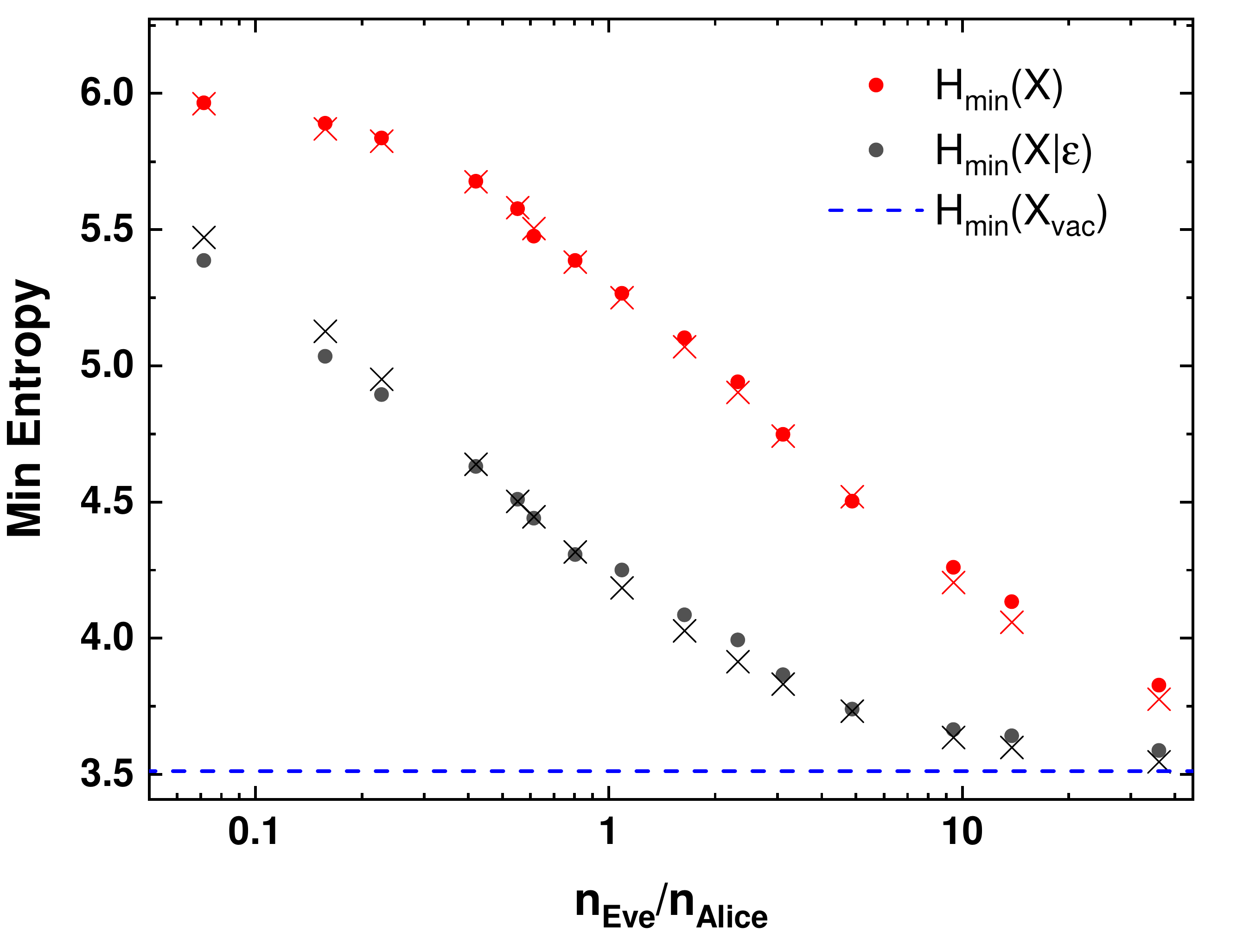}
\caption{\label{fig:ExpMinEnt} Min-entropy for the QRNG using a thermal state for different photon number ratios between the channels of Eve and Alice. Red (upper) dots represent the experimentally determined min-entropy for the bare thermal state, while black (lower) dots correspond to the reduced min-entropy in the presence of side information. Crosses give the values expected according to theory. The dashed blue line shows the expected number of guesses for a vacuum state for comparison.}
\end{figure}

\begin{figure}
\includegraphics[width=0.95\columnwidth]{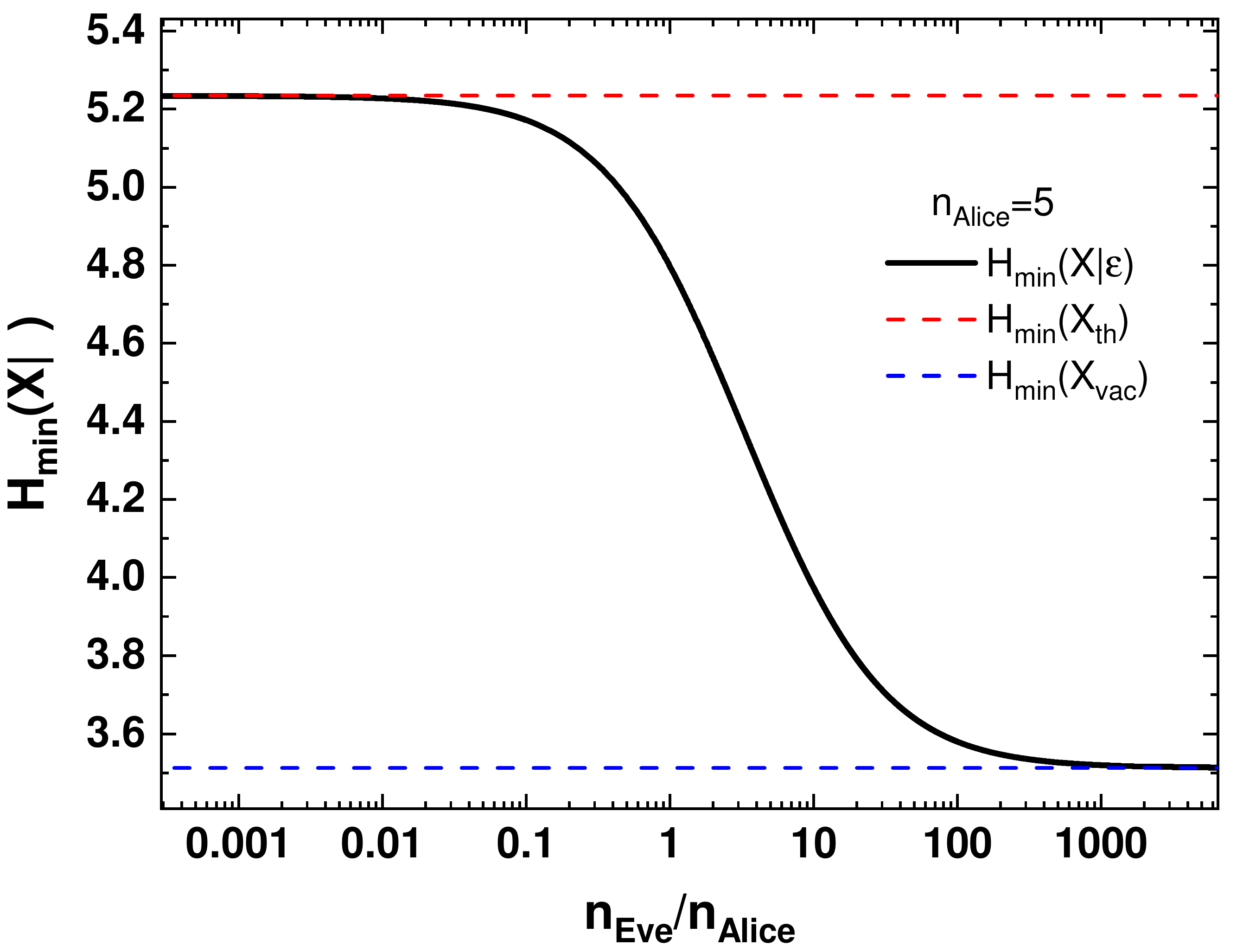}
\caption{\label{fig:Alicen5} Conditional min-entropy for a fixed photon number $n_{\text{Alice}}=5$ received by Alice. With increasing redirected photon number $n_{\text{Eve}}$, the conditional min-entropy reduces from a value corresponding to the bare thermal state to the min-entropy of a vacuum state.}
\end{figure}

We have realized this eavesdropping attack experimentally for several different distributions of the total signal photon number $n$ between the photon numbers $n_{\text{Alice}}$ and $n_{\text{Eve}}$ that reach the QRNG detection channel and the eavesdropping channel, respectively. In a real eavesdropping attempt, Eve would take care to keep $n_{\text{Alice}}$ constant to ensure that the attack is not noticed, e.g. by permanently adding a tunable attenuator in the signal beam. Instead, we deliberately choose several splitting ratios between $n_{\text{Alice}}$ and $n_{\text{Eve}}$ to investigate the physics behind the attack, whereby we increase $n_{\text{Eve}}$ while decreasing $n_{\text{Alice}}$. A list of the splitting ratios used is given in table \ref{tab:PhotonNrRatios} in the appendix. The results for the experimentally determined estimates for the min-entropy as given by the peak probability of the projected conditional Wigner function are shown in figure \ref{fig:ExpMinEnt}.
The experimental results match the theoretical predictions well. Here, red dots denote the min-entropy determined for the bare thermal signal state without taking additional side information into account. As expected, the min-entropy increases with the photon number received by Alice. The Wigner function of a thermal state is a Gaussian centered at 0 in phase space that gets broader with increasing photon number. Accordingly the probability of detecting large quadrature values increases as well, while the peak probability goes down, which is known to enhance $H_{\text{min}}(X)$ \cite{Qi2017}. For low photon numbers, $H_{\text{min}}(X)$ approaches the value of 3.51, which is the min-entropy that would be achieved for the vacuum state for the quadrature resolution given by the experimental setup. One may now compare these values to the results for $H_{\text{min}}(X|\varepsilon)$. In all cases, the conditional min-entropy when taking the side information into account is reduced compared to the bare signal state, but the magnitude of this reduction varies. For intermediate splitting ratios close to one, the reduction amounts to approximately one bit, while both for very small and very large photon numbers in Eve's detection channel the reduction becomes significantly smaller. This is obvious for small values of $n_{\text{Eve}}$. In this case, the state of the light field in Eve's part of the setup is close to the vacuum state. Due to the classical correlations between the light fields received by Eve and Alice, Eve will essentially find a light field that is proportional to the one in Alice's part of the setup, but scaled according to the relative photon numbers and with an additional vacuum contribution, which enters at the beam splitter. For small $n_{\text{Eve}}$ the latter contribution, which is uncorrelated to the signal light field, dominates and Eve gains little side information. This effect is less obvious for large $n_{\text{Eve}}$. Here, it is important to stress that in order to be able to cover a wide range of relative photon numbers, none of the photon numbers was kept constant and for all relative photon numbers, where $n_{\text{Eve}}$ is large, $n_{\text{Alice}}$ is small. Therefore, in these cases the light field received by Alice is close to the vacuum state. As the vacuum state min-entropy forms a lower bound of the min-entropy for Gaussian states, its possible reduction in the presence of side information is also limited.\\
In order to develop a deeper understanding of the interplay between $n_{\text{Eve}}$ and the amount of side information gained, we simulated eavesdropping attacks for a fixed photon number $n_{\text{Alice}}=5$ received by Alice and different splitting ratios. The results are shown in figure \ref{fig:Alicen5}.
In the limit of $n_{\text{Eve}}\rightarrow 0$, eavesdropping becomes inefficient and the conditional min-entropy matches the unconditional min-entropy of the bare thermal state, while for large values of $n_{\text{Eve}}$, $H_{\text{min}}(X|\varepsilon)$ approaches the value expected for the vacuum state. For intermediate relative photon numbers, a smooth sigmoid-like transition between both extremes takes place. This effect can be explained intuitively. After the initial signal passes the beam splitter, both Alice and Eve receive scaled versions of this original signal with added noise due to additional vacuum contributions. When Eve measures a pair of quadratures, this allows her to map her result to a phase space region describing the state of the light field in Alice's part of the setup. The width of this region determines the amount of side information available to Eve and depends on two factors. First, it is broadened by the vacuum contributions and second, due to the differing photon numbers in the two arms of the setup, the quasiprobability distribution for the arm receiving the larger photon number is spread out further in phase space as compared to the other arm. Considering first the pure signal and neglecting the additional vacuum contributions, Eve can map a finite region in phase space given by her experimental resolution to a smaller region in Alice's phase space, if she receives more photons than Alice and she can map this phase space region to a larger region in Alice's phase space, if she receives fewer photons than Alice. The former case yields more side information. In the extreme case of receiving orders of magnitude more photons than Alice, Eve can map her phase space region containing the measured quadrature to a single point in Alice's phase space. In this case, the only uncertainty about the quadrature measured by Alice arises due to the additional vacuum contributions which are not correlated to the signal and therefore set the lower bound for the conditional min-entropy Eve may reach to that of a vacuum state.\\

\section{Brute-force attacks}
So far, we have considered only the min-entropy for a given quadrature distribution, which depends solely on the largest probability that may occur. This renders it easy to determine, but it yields only limited amounts of information. Various distributions with the same peak probability, but widely varying variances will yield the same min-entropy. However, for optimal brute force attacks the expected guesswork for such distributions that share the same min-entropy may differ significantly.
\begin{figure}
\includegraphics[width=0.95\columnwidth]{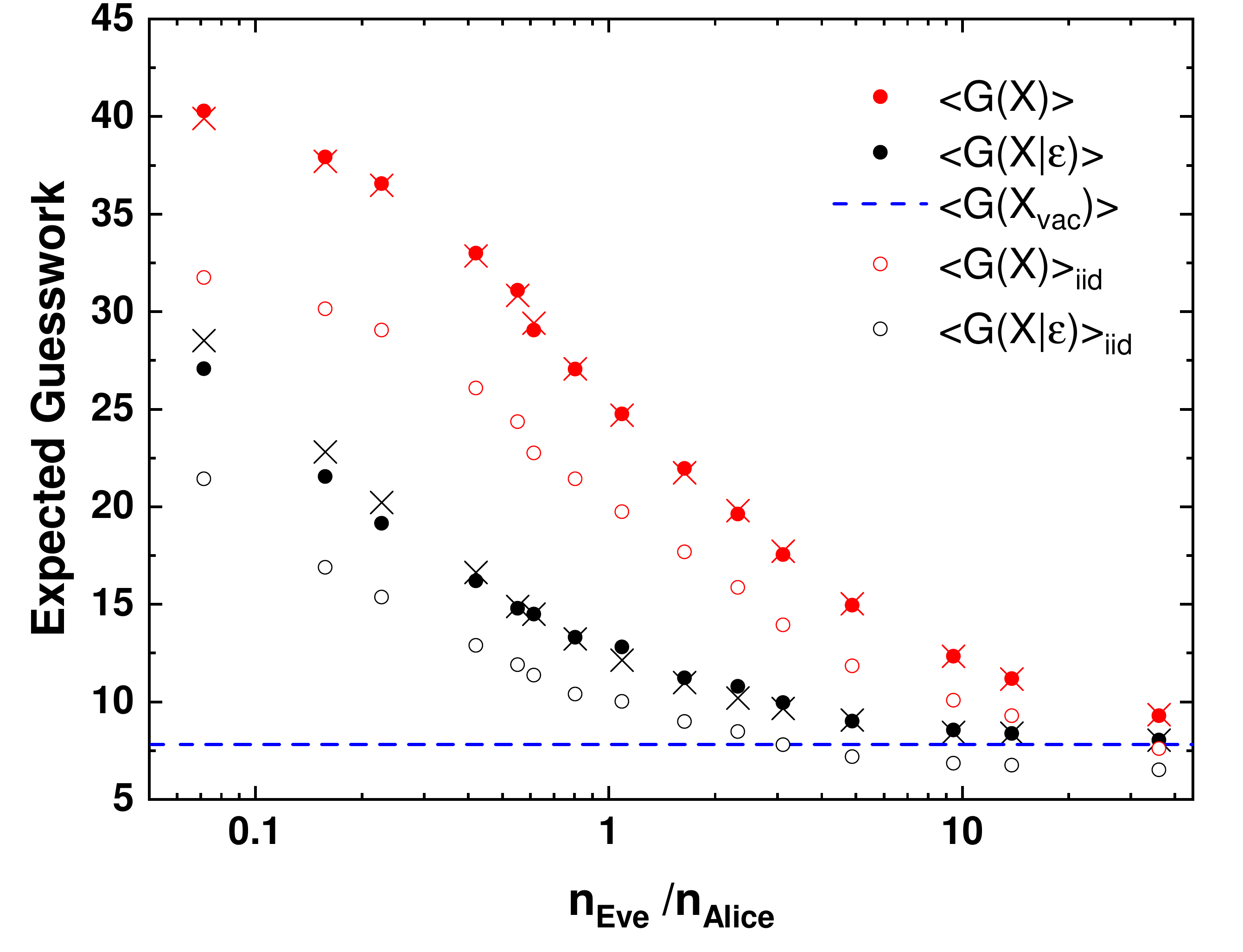}
\caption{\label{fig:ExpGuesswork} Expected guesswork for the QRNG using a thermal state for different photon number ratios between the channels of Eve and Alice. Red dots represent the experimentally determined guesswork for the bare thermal state, while black dots correspond to the reduced mean number of guesses in the presence of side information. Crosses give the values expected according to theory. The dashed blue line shows the expected number of guesses for a vacuum state for comparison. Open symbols denote the worst-case expected guessworks for a sequence of random bits with a length given by the min-entropy of the QRNG, which are too pessimistic.}
\end{figure}
Therefore, we also determined the conditional and unconditional expected guesswork for the eavesdropping attack. Results are shown in figure \ref{fig:ExpGuesswork}. The expected guesswork is clearly reduced significantly in the presence of side information. This reduction may amount to values as large as 50\% for photon number ratios $n_{\text{Eve}}/n_{\text{Alice}}$ close to and slightly below 1. A lower border is given by the value of 7.82 that corresponds to the expected guesswork for a vacuum state. This value is approximately reached for small values of $n_{\text{Alice}}$ and it should be emphasized that there is no possibility to surpass this value using the suggested eavesdropping attack. This also means that applying an eavesdropping attack of the type presented here to a vacuum state cannot yield any additional quantum side information, which is an expected result.\\Overall, the change of the expected guesswork in the presence of side information bears some resemblance to the corresponding change in min-entropy. In order to emphasize the additional value of investigating the expected guesswork, we also compare the average number of guesses required in our experiment to the worst-case prediction given by naively inserting the min-entropy for $\beta$ in eq. (\ref{eq:guesses}). These worst-case estimates are shown as open symbols both in the presence and absence of side information. These values are always lower than the mean number of guesses actually required in the experiment, which shows that the estimate for the extractable randomness given by the min-entropy is pessimistic for the brute-force attack scenario. One might also find it surprising that this number may actually drop below the expected guesswork for a vacuum state. While at first look, this result might seem to contradict the fact that one cannot reduce the min-entropy of an ideal vacuum state by gathering side information, it rather emphasizes that the min-entropy and the expected guesswork describe different physical aspects and processes. Min-entropy gives a correct worst-case estimate for Eve performing the correct guess on the first try. However, the worst-case scenario required for eq. (\ref{eq:guesses}) is different. For a binned probability distribution of known peak probability $p_{max}$, the expected number of guesses is minimized if all bins that occur with a finite probability occur with probability $p_{max}$, while all other bins do not occur at all. However, for high enough resolution, such a state is not necessarily among the physically allowed states as its variance will be smaller than the variance of the vacuum state. Such narrow distributions may still be feasible for QRNGs based on homodyning by extending the range of possible states to squeezed states, which indeed may show a smaller variance for one quadrature. However, for QRNGs based on heterodyning \cite{Avesani2018} this is not possible and the min-entropy indeed cannot be related to the expected guesswork anymore. Accordingly, we would like to emphasize that indeed the expected guesswork considers a different kind of attack than min-entropy and therefore complements it and is worthwhile to investigate as an experimental figure of merit.\\
Finally, we also demonstrate that the biased random numbers we receive from our QRNG can be turned into true random numbers. To this end, we perform two-universal hashing on a set of random numbers \cite{Frauchiger2013}, which turns 8-bit strings of biased random numbers into 4-bit strings of true random numbers and merges two consecutive numbers into a single 8-bit string of true random numbers. In order to test whether these random numbers are really independent and distributed identically, we again investigate the expected guesswork required by Eve to identify the random numbers.
\begin{figure}
\includegraphics[width=0.95\columnwidth]{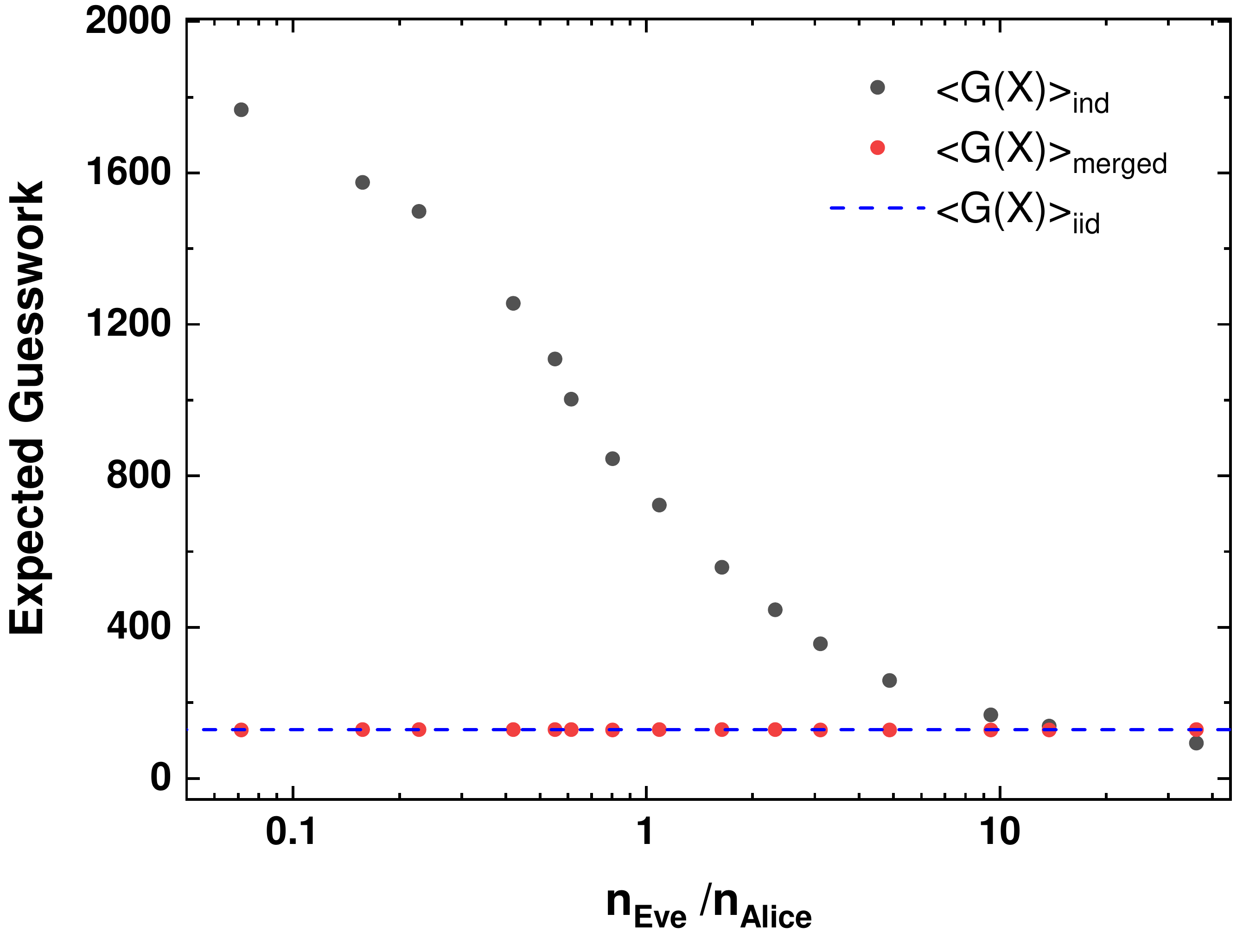}
\caption{\label{fig:MergedGuess} Expected guesswork for random numbers after performing randomness extraction and merging two consecutively measured biased random numbers into an 8-bit true random number. $\langle G(X)\rangle_{\text{ind}}$ represents the mean number of guesses required when Eve tries to guess both biased random numbers simultaneously, while $\langle G(X)\rangle_{\text{merged}}$ is the mean number of guesses needed to guess the true random number using a brute force approach. For comparison, the expected guesswork required for a true 8-bit random number is indicated by a dashed blue line.}
\end{figure}
As can be seen directly from eq.(\ref{eq:guesses}), the expected number of guesses for true 8-bit random numbers amounts to 128.5. We now determine the expected guesswork for Eve adapting two different strategies. First, $\langle G(X)\rangle_{\text{ind}}$ represents the number of guesses she needs to make when trying to guess the combination of both individual biased random numbers. To this end, Eve uses her side information gained via the eavesdropping attack and sorts all possible combinations of biased random numbers such that her first guess is the combination with the highest combined probability and sorts all combinations in order of decreasing probability. $\langle G(X)\rangle_{\text{merged}}$ instead represents a standard brute force attack, where Eve directly tries to guess the merged 8-bit random number and simply starts from bin 1 and just picks the bins in ascending order. The results are shown in figure \ref{fig:MergedGuess}. $\langle G(X)\rangle_{\text{merged}}$ always yields values close to the expected $\langle G(X)\rangle_{\text{iid}}=128.5$ guesses, which does not certify true randomness, but at least shows that trivial correlations are absent in the data. More interestingly, in most cases $\langle G(X)\rangle_{\text{ind}}$ is significantly enhanced up to values of 1800 guesses compared to the value expected for true random numbers. This shows that although Eve has access to some side information about the biased random numbers, this is not enough to also yield side information about the merged random numbers after performing randomness extraction. Only for the largest ratio of $n_{\text{Eve}}/n_{\text{Alice}}$, the mean number of guesses required by Eve is slightly below $\langle G(X)\rangle_{\text{iid}}$ with 93.2 guesses. This is remarkable because judging from $H_{\text{min}}(X|\varepsilon)$ alone, the extractable randomness per measurement took values below 4 bits per measurement already starting from ratios of $n_{\text{Eve}}/n_{\text{Alice}}\geq 2$. It is therefore not trivial that performing randomness extraction for this range of values results in an expected guesswork that would still be in line with true 8-bit random numbers.

\section{Discussion}
We have experimentally investigated a trusted QRNG based on homodyning a thermal light field subject to a heterodyne eavesdropping attack. The presence of side information reduces both the min-entropy and the expected guesswork towards the values that may be obtained by using a vacuum state instead of a thermal light field. This result has immediate implications for QRNGs based on continuous variables. QRNGs based on thermal light have been shown to offer enhanced min-entropy compared to QRNGs sampling vacuum, when using the same experimental resolution \cite{Qi2017}, but they either need to be operated as trusted QRNGs or used in applications where security of the random numbers is not relevant. Otherwise, eavesdropping attacks may reduce the performance of the thermal light QRNG such, that it does not offer any advantages compared to sampling vacuum. At this point, it is worthwhile to emphasize again the assumptions that our experimental study is based on. First, the lower bounds derived rely on the assumption that the QRNG is trusted in the sense that the prepared signal state indeed is a thermal state, uncorrelated with respect to the eavesdropper. A malicious QRNG manufacturer might as well prepare a two-mode squeezed vacuum state, which will also look thermal to Alice and would result in lower min-entropy \cite{Marangon2017} or single-mode squeezed light. It should be noted that a QRNG based on heterodyning instead of homodyning would be less susceptible to such attacks. Also, the manufacturer might mimic the thermal state by preparing coherent states of different amplitude and phase, which would also directly yield side information, which reduces the min-entropy of the signal to the min-entropy of a vacuum state. Along the same lines, the deconvolution procedure used by Eve to calculate the conditional Wigner function from the conditional Husimi function requires the assumption that the signal light field is in some kind of Gaussian state. Otherwise the deconvolution procedure will be more sophisticated and not necessarily unambiguous.\\We also implicitly assume that the repetition rate of the local oscillator is low enough to avoid memory attacks. As is well known, a thermal light field may be described as a random walk in phase space, where the instantaneous light field at any time is given by a coherent state, which dephases on a timescale given by the coherence time of the light field. This means that two consecutive quadrature measurements are not necessarily independent of each other. For time delays shorter than the coherence time of the light field, the measured values will still be correlated, which will also provide side information. This problem may always be avoided by choosing a delay between consecutive measurements that is long compared to the coherence time of the light field. In our case, the coherence time of the light field amounts to about 750\,fs, while the delay between local oscillator pulses amounts to 13.3\,ns, so the individual measurements can be considered as independent of each other.\\
While the present study corresponds to a proof of principle demonstration, it would be very interesting for future studies to repeat such an eavesdropping attack on a QRNG that performs heterodyning on a thermal state and investigate the expected guesswork for such a high-performance device. Finally, we would like to emphasize that from a spectroscopic point of view the central figure of merit in our manuscript is the expected guesswork and that it represents a quantity worth studying in experiments. It is obvious that the expected guesswork is of limited interest with respect to the security of QRNGs as no reasonable protocol using private random numbers would allow for a brute-force attack. However, for example it is possible to map the problem of eavesdropping on a continuous variable QRNG to the problem of performing time-resolved quantum state tomography on a classical state of the light field in the absence of a fixed phase reference between the LO and the signal. In this problem, Eve's measurement corresponds to an ancilla measurement used to guess the instantaneous phase and amplitude of the signal light field, while Alice may use this information to perform quantum state tomography or conditional measurements of the signal at different times relative to Eve. In this case, the expected guesswork yields information that may be used to optimize the state reconstruction process.


\begin{acknowledgments}
We gratefully acknowledge funding by the Deutsche Forschungsgemeinschaft (DFG, German Research Foundation) via project number 231447078 – TRR 142, project A4.
\end{acknowledgments}

\appendix

\section{Detector properties}
\begin{figure*}[!ht]
\includegraphics[width=0.75\textwidth]{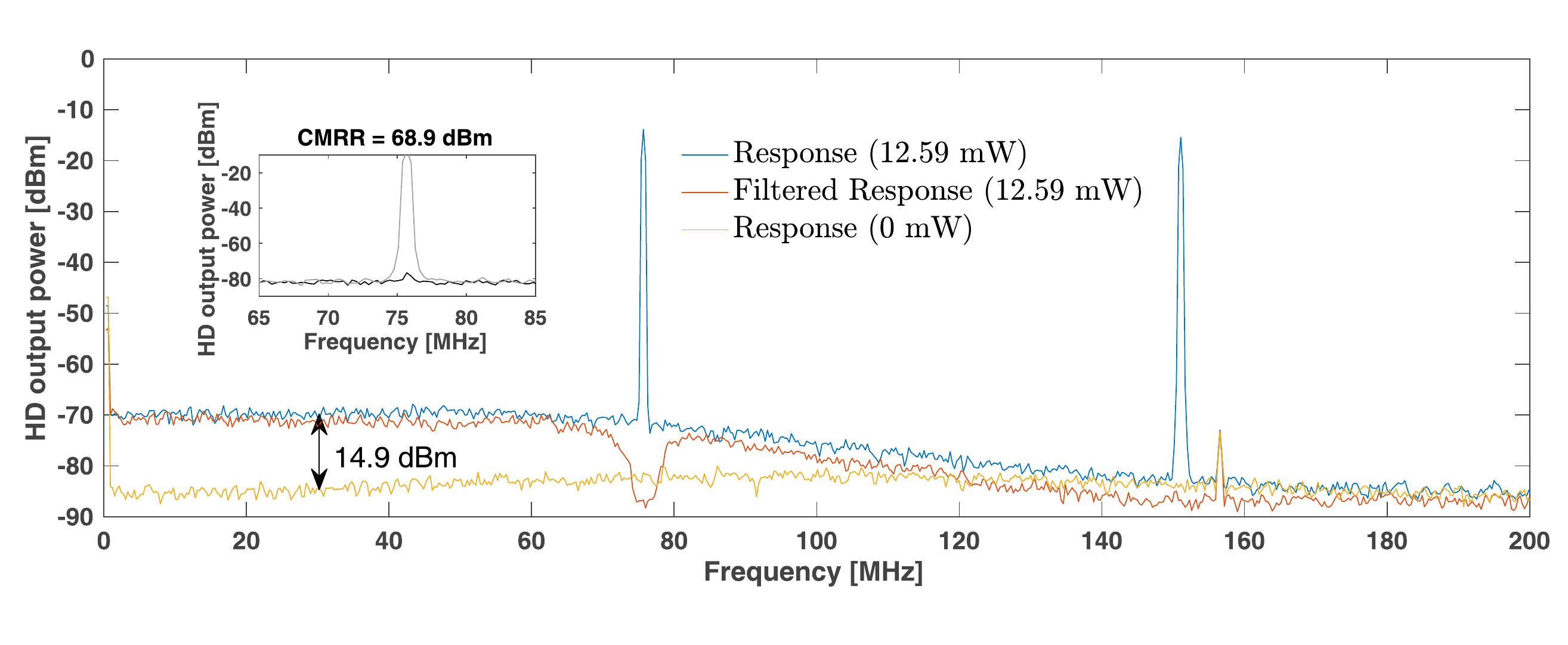}
\caption{\label{fig:PowerSpec} Frequency response of the homodyne detector: The lowest detector output corresponds to the absence of any light field at the detector. When applying a local oscillator with a power of 12.59\,mW, the signal level rises and characteristic peaks at the laser repetition frequency of 75.39\,MHz and its multiples arise. These are removed by adding notch filters. In this case, the low-frequency shot-noise clearance amounts to approximately\,15 dBm and we reach a common-mode rejection ratio of about 69\,dBm.}
\end{figure*}

In the main text, we presented the central figures of merit for our detectors and outlined the calibration procedure. These can be derived from the frequency response of the detector. One example comparing the response at high LO powers and for no signal at all is shown in figure \ref{fig:PowerSpec}. 

\begin{figure}[!ht]
\includegraphics[width=0.95\columnwidth]{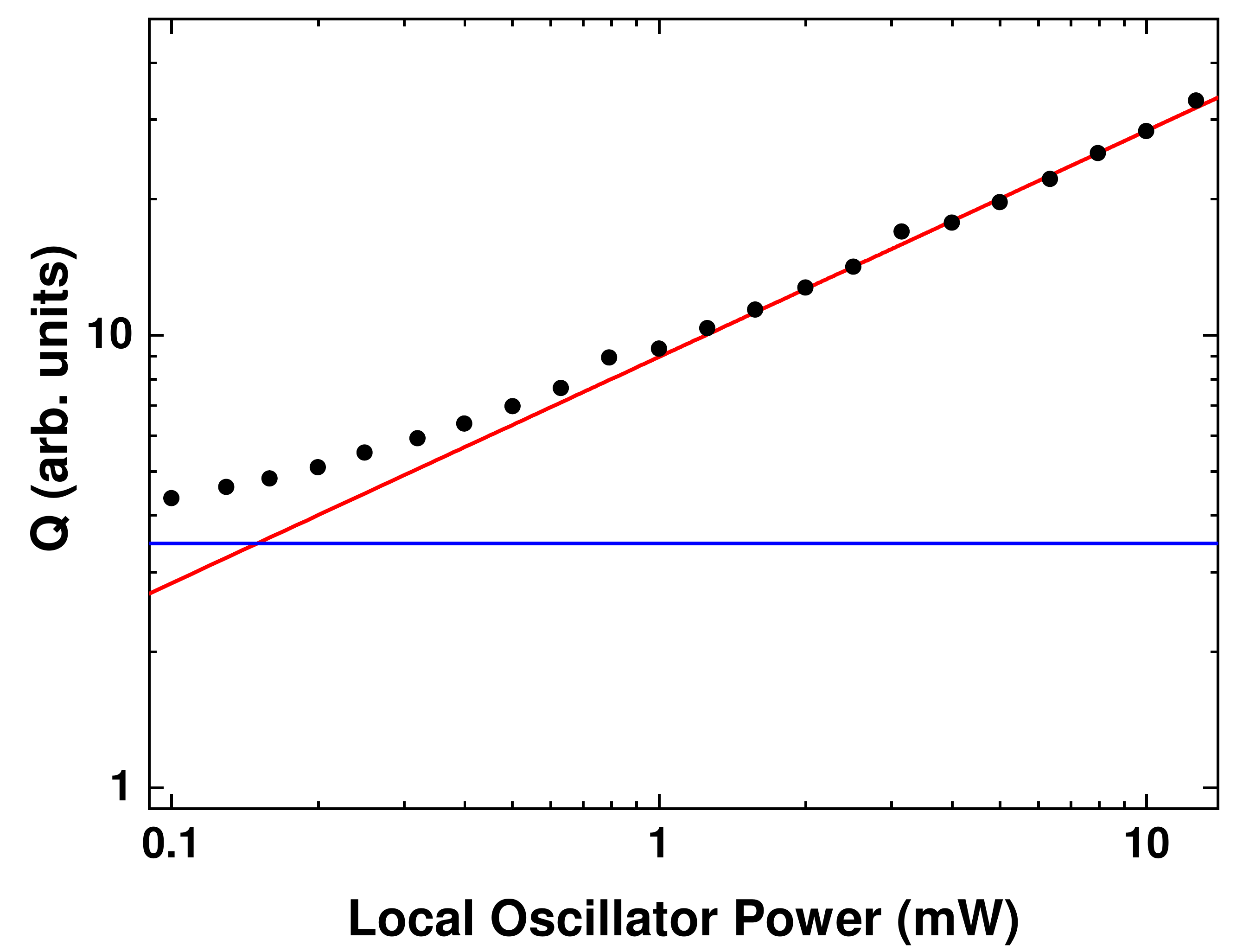}
\caption{\label{fig:ShotNoise} Measured standard deviation of the unnormalized vacuum state quadrature values for different LO powers. The constant blue line and the red line denote the electronic detector noise and the shot noise limit, respectively.}
\end{figure} 

A low-frequency shot-noise clearance of up to 15\,dBm may be reached. The results presented in the main text were achieved using a smaller LO power of 5\,mW. While this results in a slightly reduced shot-noise clearance, it also results in a larger dynamic range, which is beneficial for thermal light, which shows a large variance of quadrature values. Still, it is important to verify that the detector is still operating in the shot-noise limit at this reduced LO power. We demonstrate this by measuring the standard deviation of the vacuum state quadratures measured for several LO powers as shown in figure \ref{fig:ShotNoise}. Here, the dots represent the experimental results, the constant blue line denotes the basic electronic detector noise and the red line represents the shot noise limit that scales as the square root of the LO power. As can be seen clearly, the detector starts to operate in the shot noise-limited region for LO powers larger than approximately 1\,mW.

\section{Randomness extraction and statistical tests}
The focus of our manuscript is on the physical implementation of the attack instead of optimizing the performance of a QRNG. Accordingly, we used only a strongly simplified toy version of standard randomness extraction procedures. We follow the procedure given in \cite{Frauchiger2013}. Instead of using Toeplitz matrices, we use simple random matrices of size 32x16 containing 16 bit integers and a fixed bit extraction rate of 4 bit per measurement. Details on this procedure can be found in appendix B. In order to test our implementation of the quantum random number generator, we also performed statistical tests on the generated random numbers. We used the NIST suite to test them \cite{Bassham2010}. 
\begin{table}[!h]
\begin{tabular}{ccc}
\hline
   Test name & P-Value & Result\\
	\hline
  Frequency & 0.834 & PASSED \\
	BlockFrequency & 0.067 & PASSED\\
	CumulativeSums & 0.478 & PASSED\\
	Runs & 0.689 & PASSED\\
	LongestRun & 0.834 & PASSED\\
	Rank & 0.637 & PASSED\\
	FFT & 0.834 & PASSED\\
	NonOverlappingTemplate & 0.421 & PASSED\\
	OverlappingTemplate & 0.312 & PASSED\\
	Universal & 0.789 & PASSED\\
	ApproximateEntropy & 0.834 & PASSED\\
	RandomExcursions & 0.477 & PASSED\\
	RandomExcursionsVariant & 0.456 & PASSED\\
	Serial & 0.739 & PASSED\\
	LinearComplexity & 0.689 & PASSED\\	
	\hline
	\hline
\end{tabular}
\caption{\label{tab:NISTTable} Result of the NIST test suite applied to the random numbers extracted from the continuous variable quantum random number generator. Whenever there are several tests in a category, the mean P-value is reported.}
\end{table}
All tests were passed as shown in table \ref{tab:NISTTable}. Passing these tests does by no means certify that the numbers are random, but demonstrates that several types of patterns are absent. 

\section{Photon Number Ratios}
Within the manuscript several different ratios between $n_{\text{Eve}}$ and $n_{\text{Alice}}$ have been investigated. For completeness, table \ref{tab:PhotonNrRatios} below shows the exact photon numbers used in the experiment.

\begin{table}[!h]
\begin{tabular}{ccc}
\hline
   $n_{\text{Eve}}$ & $n_{\text{Alice}}$ & $n_{\text{Eve}}/n_{\text{Alice}}$\\
	\hline
  1.04 & 14.60 & 0.07 \\
	2.01 & 12.78 & 0.16\\
	2.73 & 11.98 & 0.23\\
	4.06 & 9.64 & 0.42\\
	4.64 & 8.40 & 0.55\\
	4.59 & 7.49 & 0.61\\
	5.02 & 6.24 & 0.80\\
	5.58 & 5.12 & 1.09\\
	6.34 & 3.87 & 1.64\\
	6.89 & 2.97 & 2.32\\
	7.09 & 2.28 & 3.11\\
	7.53 & 1.54 & 4.89\\
	7.70 & 0.82 & 9.45\\
	7.90 & 0.57 & 13.81\\
	8.05 & 0.22 & 36.10\\
	\hline
	\hline
\end{tabular}
\caption{\label{tab:PhotonNrRatios} Photon numbers used for the experiments in the manuscript.}
\end{table}

\newpage


%


\end{document}